\documentstyle[twoside,fleqn,epsf,my_espcrc2]{article}

\global\arraycolsep=1.5pt
\def\spose#1{\hbox to 0pt{#1\hss}}
\def\lsim{\mathrel{\spose{\lower 3pt\hbox{$\mathchar"218$}}
 \raise 2.0pt\hbox{$\mathchar"13C$}}}
\def\gsim{\mathrel{\spose{\lower 3pt\hbox{$\mathchar"218$}}
 \raise 2.0pt\hbox{$\mathchar"13E$}}}

\begin{document}

\begin{titlepage}

\begin{flushright}
CERN-TH/97-19\\
hep-ph/9702310
\end{flushright}

\vspace{2cm}

\begin{center}
\Large\bf Theory of Inclusive B Decays
\end{center}

\vspace{2cm}

\begin{center}
Matthias Neubert\\
{\sl Theory Division, CERN, CH-1211 Geneva 23, Switzerland}
\end{center}

\vspace{1cm}

\centerline{\bf Abstract}
\medskip
\centerline{\parbox{12cm}{
We present the theory of inclusive decays of hadrons containing a
heavy quark and discuss its most important applications to the decays
of $B$ mesons. We also review the theoretical understanding of the
hadronic parameters $\lambda_1$ and $\lambda_2$ (or $\mu_\pi^2$ and
$\mu_G^2$) entering the heavy-quark expansion.
}}

\vspace{1.5cm}

\begin{center}
Invited talk presented at the\\
4th KEK Topical Conference on Flavour Physics\\
KEK, Japan, 29-31 October 1996
\end{center}

\vspace{2cm}

\noindent
CERN-TH/97-19\\
February 1997
\vfil

\end{titlepage}

\thispagestyle{empty}
\vbox{}
\newpage

\setcounter{page}{1}


\title{Theory of Inclusive B Decays}

\author{Matthias Neubert\\[0.3cm]
Theory Division, CERN\\
CH-1211 Geneva 23, Switzerland}


\begin{abstract}
We present the theory of inclusive decays of hadrons containing a
heavy quark and discuss its most important applications to the decays
of $B$ mesons. We also review the theoretical understanding of the
hadronic parameters $\lambda_1$ and $\lambda_2$ (or $\mu_\pi^2$ and
$\mu_G^2$) entering the heavy-quark expansion.
\end{abstract}

\maketitle

\section{INTRODUCTION}

Hadronic bound states of a heavy quark with light constituents
(quarks, antiquarks and gluons) are characterized by a large
separation of mass scales: the heavy-quark mass $m_Q$ is much larger
than the mass scale $\Lambda_{\rm QCD}$ associated with the light
degrees of freedom. Equivalently, the Compton wave length of the
heavy quark ($\lambda_Q\sim 1/m_Q$) is much smaller than the size of
the hadron containing it ($R_{\rm had}\sim 1/\Lambda_{\rm QCD}$). In
such a situation, it is appropriate to separate the physics
associated with these two scales, so that all dependence on the
heavy-quark mass becomes explicit. The framework in which to perform
this separation is the heavy-quark (or $1/m_Q$) expansion
\cite{review}--\cite{Grorev}, which is a specific realization of the
operator product expansion (OPE) \cite{Wils,Zimm}.

There are (at least) two important reasons why it is desirable to
separate short- and long-distance phenomena: A technical reason is
that after this separation we can calculate a big portion of the
relevant physics (i.e.\ all short-distance effects) using
perturbation theory and renormalization-group techniques; in
particular, in this way we are able to control all logarithmic
dependence on the heavy-quark mass. An important physical reason is
that, after the short-distance physics has been separated, the
long-distance physics may simplify due to the realization of
approximate symmetries, which relate the long-distance properties of
many observables to a small number of hadronic matrix elements. The
second point is particularly exciting, since it allows us to make
statements beyond the range of applicability of perturbation theory.

In our case, an approximate spin--flavour symmetry is realized in
systems in which a single heavy quark interacts with light degrees of
freedom by the exchange of soft gluons \cite{Shu1}--\cite{Isgu}. The
origin of this symmetry is not difficult to understand. In a heavy
hadron, the heavy quark is surrounded by a most complicated, strongly
interacting cloud of light quarks, antiquarks and gluons. However,
the fact that the Compton wavelength of the heavy quark is much
smaller than the size of the hadron leads to simplifications. To
resolve the quantum numbers of the heavy quark would require a hard
probe; the soft gluons exchanged between the heavy quark and the
light constituents can only resolve distances much larger than
$1/m_Q$. Therefore, the light degrees of freedom are blind to the
flavour (mass) and spin orientation of the heavy quark. They
experience only its colour field, which extends over large distances
because of confinement. In the rest frame of the heavy quark, it is
in fact only the electric colour field that is important;
relativistic effects such as colour magnetism vanish as
$m_Q\to\infty$. Since the heavy-quark spin participates in
interactions only through such relativistic effects, it decouples. It
follows that, in the limit $m_Q\to\infty$, hadronic systems which
differ only in the flavour or spin quantum numbers of their heavy
quark have the same configuration of their light degrees of freedom.
Although this observation still does not allow us to calculate what
this configuration is, it provides relations between the properties
of such particles as the heavy mesons $B$, $D$, $B^*$ and $D^*$, or
the heavy baryons $\Lambda_b$ and $\Lambda_c$.

Heavy-quark symmetry is an approximate symmetry, and corrections
arise since the quark masses are not infinitely heavy. Nevertheless,
results derived on the basis of heavy-quark symmetry are
model-independent consequences of QCD in a well-defined limit. The
symmetry-breaking corrections can, at least in principle, be studied
in a systematic way. A convenient framework for analysing these
corrections is given by the heavy-quark effective theory (HQET)
\cite{Geor}, which provides a systematic expansion around the limit
$m_Q\to\infty$. In the HQET, a heavy quark inside a hadron moving
with velocity $v$ is described by a velocity-dependent field
$h_v$ subject to the constraint $\rlap/v\,h_v=h_v$. This field is
related to the original heavy-quark field by a phase redefinition, so
that it carries the ``residual momentum'' $k=p_Q-m_Q v$, which
characterizes the interactions of the heavy quark with gluons. The
effective Lagrangian of the HQET is \cite{Geor}--\cite{Mann}
\begin{eqnarray}
   {\cal L}_{\rm eff} &=& \bar h_v\,i v\!\cdot\!D\,h_v
    + \frac{C_{\rm kin}}{2 m_Q}\,\bar h_v (i D_\perp)^2 h_v
    \nonumber\\
   &&\mbox{}+ \frac{C_{\rm mag}\,g_s}{4 m_Q}\,
    \bar h_v\sigma_{\mu\nu} G^{\mu\nu} h_v + O(1/m_Q^2) \,,
\label{Leff}
\end{eqnarray}
where $D_\perp^\mu=D^\mu-(v\cdot D)\,v^\mu$ contains the components
of the gauge-covariant derivative orthogonal to the velocity, and
$g_s G^{\mu\nu}=i[D^\mu,D^\nu]$ is the gluon field-strength tensor.
The leading term in the effective Lagrangian, which gives rise to the
Feynman rules of the HQET, is invariant under a global $SU(2n_h)$
spin--flavour symmetry group, where $n_h$ is the number of
heavy-quark flavours. This symmetry is explicitly broken by the
higher-dimensional operators arising at order $1/m_Q$, whose origin
is most transparent in the rest frame of the heavy hadron: the first
operator corresponds to (minus) the kinetic energy resulting from the
motion of the heavy quark inside the hadron (in the rest frame, $(i
D_\perp)^2$ is the operator for $-{\bf k}^2$), and the second
operator describes the chromo-magnetic interaction of the heavy-quark
spin with the gluon field. The coefficients $C_{\rm kin}$ and $C_{\rm
mag}$ result from short-distance effects and, in general, depend on
the scale at which the operators are renormalized. At the tree level,
$C_{\rm kin}=C_{\rm mag}=1$.

At this point it is instructive to recall a more familiar example of
how approximate symmetries relate the long-distance properties of
several observables. The strong interactions of pions are severely
constrained by the approximate chiral symmetry of QCD. In a certain
kinematic regime, where the momenta of the pions are much less than
1\/GeV (the scale of chiral-symmetry breaking), the long-distance
physics of scattering amplitudes is encoded in a few ``reduced matrix
elements'', such as the pion decay constant. An effective low-energy
theory called chiral perturbation theory provides a systematic
expansion of scattering amplitudes in powers of the pion momenta, and
thus helps to derive the relations between different scattering
amplitudes imposed by chiral symmetry \cite{Leut}. A similar
situation holds for the case of heavy quarks. Heavy-quark symmetry
implies that, in the limit where $m_Q\gg\Lambda_{\rm QCD}$, the
long-distance properties of several observables is encoded in few
hadronic parameters, which can be defined in terms of operator matrix
elements in the HQET.

In particular, the forward matrix elements of the two dimension-five
operators in (\ref{Leff}),
\begin{eqnarray}
   O_{\rm kin} &=& - \bar h_v (i D_\perp)^2 h_v \,, \nonumber\\
   O_{\rm mag} &=& \frac{g_s}{2}\,\bar h_v\sigma_{\mu\nu}
    G^{\mu\nu} h_v \,,
\end{eqnarray}
play a most significant role in many applications of the HQET. They
appear, for instance, in the spectroscopy of heavy hadrons and in the
description of inclusive weak decays \cite{review}. For the $B$
meson, we define two parameters $\lambda_1$ and $\lambda_2$ (or,
equivalently, $\mu_\pi^2$ and $\mu_G^2$) by \cite{FaNe}
\begin{eqnarray}
   -\lambda_1 &=& \mu_\pi^2
    = \frac{1}{2 m_B}\,\langle B|\,O_{\rm kin}\,|B\rangle \,,
    \nonumber\\
   3\lambda_2 &=& \mu_G^2
    = {1\over 2 m_B}\,\langle B|\,O_{\rm mag}\,|B\rangle \,.
\end{eqnarray}
Whereas $\lambda_2$ is directly related to the mass splitting between
vector and pseudoscalar mesons through
\begin{equation}
   m_{B^*}^2 - m_B^2 = 4\lambda_2 + O(1/m_b) \,,
\label{split}
\end{equation}
the parameter $\lambda_1$ cannot be determined from hadron
spectroscopy. It is, however, related to the difference of the pole
masses of two heavy quarks through
\begin{eqnarray}
   m_b - m_c &=& (\overline{m}_B - \overline{m}_D) \nonumber\\
   &&\mbox{}+ \lambda_1 \left( \frac{1}{2\overline{m}_B}
    - \frac{1}{2\overline{m}_D} \right) + \dots \,.
\label{mQdif}
\end{eqnarray}
Here $\overline{m}_B=\frac 14(m_B+3m_{B^*})$ and
$\overline{m}_D=\frac 14(m_D+3m_{D^*})$ denote the spin-averaged
meson masses, defined such that they do not receive a contribution
from the chromo-magnetic interaction. In the following section, we
discuss the theoretical status of the parameters $\lambda_1$ and
$\lambda_2$ and their properties under renormalization.

\section{OPERATOR RENORMALIZATION AND THE VIRIAL THEOREM}

In quantum field theory, local composite operators such as $O_{\rm
kin}$ and $O_{\rm mag}$ require renormalization to be well defined.
The matrix elements of these operators (i.e.\ $\lambda_1$ and
$\lambda_2$) depend on the subtraction scheme and on the
renormalization scale $\mu$ introduced in the process of removing
ultraviolet divergences. This dependence is cancelled by an opposite
scale dependence of the Wilson coefficient functions $C_{\rm kin}$
and $C_{\rm mag}$ in the effective Lagrangian (\ref{Leff}). The most
common regularization scheme in QCD is dimensional regularization
\cite{tHo2,Boll} with modified minimal subtraction
($\overline{\mbox{MS}}$ scheme \cite{MSbar}). Ultraviolet divergences
are regulated by working in $d=4-2\epsilon$ space--time dimensions,
and are subtracted by removing the poles in $1/\epsilon$. The scale
dependence of the Wilson coefficients is governed by
renormalization-group equations of the form
\begin{equation}
   \mu\,\frac{\mbox{d}}{\mbox{d}\mu}\,C(\mu) = \gamma\,C(\mu) \,,
\label{RGE}
\end{equation}
where $\gamma$ is called the anomalous dimension.

The anomalous dimension of the kinetic operator $O_{\rm kin}$
vanishes to all orders in perturbation theory, $\gamma_{\rm
kin}\equiv 0$. This is a consequence of the reparametrization
invariance of the HQET, i.e.\ and invariance under infinitesimal
changes of the velocity $v$ used in the construction of the effective
Lagrangian \cite{LuMa}--\cite{Fink}. It follows that in dimensional
regularization $C_{\rm kin}\equiv 1$. The anomalous dimension of the
chromo-magnetic operator does not vanish, however. It has been known
at the one-loop order since many years \cite{EiH2,FGL}, but only very
recently the two-loop coefficient has been calculated
\cite{ABM,GrCz}. For $N_c=3$ colours, the result is
\begin{equation}
   \gamma_{\rm mag} = \frac{3\alpha_s}{2\pi}
   + \!\Big( 17 - \frac{13}{6}\,n_f \Big)\!
   \left( \frac{\alpha_s}{2\pi} \right)^2\! + O(\alpha_s^3) \,,
\end{equation}
where $n_f$ is the number of light-quark flavours. Given this result,
together with the one-loop matching condition \cite{EiH2}
\begin{equation}
   C_{\rm mag}(m_Q) = 1 + \frac{13}{6}\,
   \frac{\alpha_s(m_Q)}{\pi} + O(\alpha_s^2) \,,
\end{equation}
one can work out the next-to-leading order expression for $C_{\rm
mag}(\mu)$, starting from the general solution to (\ref{RGE})
\cite{MSbar,Flor}:
\begin{equation}
   C(\mu) = C(m_Q)\,\exp\int\limits_{\alpha_s(m_Q)}^{\alpha_s(\mu)}
   \!\mbox{d}\alpha_s\,\frac{\gamma(\alpha_s)}{\beta(\alpha_s)} \,,
\end{equation}
where $\beta(\alpha_s)=\mbox{d}\alpha_s(\mu)/\mbox{d}\ln\mu$ is the
$\beta$ function. The product of the Wilson coefficient $C_{\rm
mag}(\mu)$ and the scale-dependent matrix element $\lambda_2(\mu)$ is
renormalization-group invariant. Hence, in the presence of
renormalization effects, what can be determined from the
vector--pseudoscalar mass splitting in (\ref{split}) is the
combination
\begin{equation}
   \lambda_2 \equiv C_{\rm mag}(m_b)\,\lambda_2(m_b)
   \approx 0.12\,\mbox{GeV}^2 \,.
\label{lam2val}
\end{equation}
The numerical value is specific for the $B$ system, since according
to our definition the renormalized parameter $\lambda_2$ depends
logarithmically on the $b$-quark mass.

Unfortunately, the question about the value of the kinetic-energy
parameter $\lambda_1$ is more difficult to answer, even from a
conceptual point of view. We have already mentioned that
spectroscopic relations involving $\lambda_1$, such as (\ref{mQdif}),
depend on the heavy-quark pole masses, which are not physical
parameters. In the remainder of this section, we shall argue that
indeed the parameter $\lambda_1$ is not physical, but has an
intrinsic ambiguity of order $\Lambda_{\rm QCD}^2$. (However, the
difference of the values of $\lambda_1$ in two different hadrons is a
physical parameter and can be extracted from spectroscopy.) This
means that $\lambda_1$ must be defined in a non-perturbative way; in
other words, different definitions cannot be simply related to each
other using perturbation theory. The wide spread in the theoretical
predictions for the parameter $\lambda_1$, shown in
Tab.~\ref{tab:kinetic}, is partially a reflection of this problem.
Future efforts should concentrate on understanding better the
relations between the various definitions underlying these estimates.

\begin{table}
\caption{Theoretical estimates of the parameter $\lambda_1$ (QCDSR:
QCD sum rules, HQSR: heavy-quark sum rules, Exp.: experimental data
on inclusive decays, QM: quark models)}
\vspace{0.4cm}
\begin{tabular}{|l|l|l|}\hline
Reference \rule[-0.25cm]{0cm}{0.7cm} & Method &
 $-\lambda_1$ [GeV$^2$] \\
\hline
Eletsky, Shuryak \cite{ElSh} \hspace{-3mm} \rule{0cm}{0.4cm} &
 QCDSR & $0.18\pm 0.06$ \\
Ball, Braun \cite{BaBr} & QCDSR & $0.52\pm 0.12$ \\
Neubert \cite{l1sr} & QCDSR & $0.10\pm 0.05$ \\[0.06cm]
\hline
Gim\'enez et al.\ \cite{GiMS} \rule{0cm}{0.4cm} & Lattice &
 $-0.09\pm 0.14\!$ \\[0.06cm]
\hline
Bigi et al.\ \cite{BSUV} \rule{0cm}{0.4cm} & HQSR & $>0.36$
 \\[0.06cm]
\hline
Gremm et al.\ \cite{GKLW} \rule{0cm}{0.4cm} & Exp. &
 $0.19\pm 0.10$ \\
Falk et al.\ \cite{FLS2} & Exp. & $\approx 0.1$ \\
Chernyak \cite{Cher} & Exp. & $0.14\pm 0.03$ \\[0.06cm]
\hline
Hwang et al.\ \cite{Hwan} \rule{0cm}{0.4cm} & QM &
 $0.5\pm 0.1$ \\
De Fazio\ \cite{Fazi} & QM & $0.66\pm 0.13$ \\[0.06cm]
\hline
\end{tabular}
\label{tab:kinetic}
\end{table}

The problem of the ambiguity in the value of the heavy-quark kinetic
energy is closely related to the well-known ambiguity in the
definition of the pole mass of a heavy quark \cite{BBren,Bigiren}.
The reason lies in the divergent behaviour of perturbative expansions
in large orders, which is associated with the existence of
singularities along the real axis in the Borel plane, the so-called
renormalons \cite{tHof}--\cite{Muel}. When the pole mass $m_Q$ is
related to a short-distance mass (such as the running mass in the
$\overline{\mbox{MS}}$ scheme), which is a well-defined quantity in
perturbation theory, then the corresponding perturbation series
\begin{eqnarray}
   m_Q &=& m_Q^{\rm SD}\,\Big\{ 1 + c_1\,\alpha_s(m_Q)
    + c_2\,\alpha_s^2(m_Q) + \dots \nonumber\\
   &&\qquad \mbox{}+ c_n\,\alpha_s^n(m_Q) + \dots \Big\} \,,
\end{eqnarray}
contains numerical coefficients $c_n$ that grow as $n!$ for large
$n$, rendering the series divergent and not Borel summable. The best
one can achieve is to truncate the series at the minimal term, but
this leads to an unavoidable arbitrariness of order $\Delta
m_Q\sim\Lambda_{\rm QCD}$ (the size of the minimal term). This
observation, which at first sight seems a serious problem, should not
come as a surprise. We know that because of confinement quarks do not
appear as physical states in nature. Hence, there is no unique way to
define their on-shell properties such as a pole mass. In view of
this, it is actually remarkable that QCD perturbation theory
``knows'' about its incompleteness and indicates, through the
appearance of renormalon singularities, the presence of
non-perturbative effects. We must first specify a scheme how to
truncate the QCD perturbation series before non-perturbative
parameters such as $m_Q$ and $\lambda_1$ become well-defined
quantities. The actual values of these parameters will depend on this
scheme.

In the difference of the pole masses on the left-hand side in
(\ref{mQdif}), the leading renormalon ambiguity of order
$\Lambda_{\rm QCD}$ cancels. However, there remains a residual
ambiguity of order $\Lambda_{\rm QCD}^2/m_Q$ \cite{explo}. This point
was unclear for some time, since the corresponding renormalon
singularity does not appear in the so-called bubble approximation, in
which most explicit calculations in the Borel plane are performed
\cite{BBren,MNS}. However, it is now known that this singularity must
be present beyond the bubble approximation, and the ambiguity in the
difference $m_b-m_c$ is then absorbed by a corresponding ambiguity of
order $\Lambda_{\rm QCD}^2$ in the parameter $\lambda_1$ in
(\ref{mQdif}) \cite{explo}.

It is instructive to consider the same problem from a different point
of view, by studying the properties of the kinetic operator under
renormalization in regularization schemes with a dimensionful cutoff
parameter $\mu$. In such schemes, the operator $O_{\rm kin}$ can mix
with the lower-dimensional operator $\bar h_v h_v$ (the
``identity''), because the two operators have the same quantum
numbers. If such a mixing is present, it leads to an additive
contribution to the parameter $\lambda_1$ of the form
$\mu^2\,C[\alpha_s(\mu)]$, which one would like to subtract in order
to define a renormalized parameter. The coefficient $C$ can be
calculated order by order in an expansion in the small coupling
constant $\alpha_s(\mu)$, and it appears at first sight that the
quadratically divergent term could be subtracted using perturbation
theory. This impression is erroneous, however, because $C$ may
contain non-perturbative contributions of the form
$\exp[-8\pi/\beta_0\alpha_s(\mu)]=(\Lambda_{\rm QCD}/\mu)^2$, which
cannot be controlled in perturbation theory \cite{MMS}. Such terms
contribute an amount of order $\Lambda_{\rm QCD}^2$ to the parameter
$\lambda_1$, which is of the same order as the renormalized parameter
itself. Hence, if the kinetic operator mixes with the identity, it is
necessary that the quadratically divergent contribution to
$\lambda_1$ be subtracted in a non-perturbative way, and hence the
heavy-quark kinetic energy by itself is not directly a physical
quantity.

The question whether there is a mixing of the kinetic energy with the
identity, and whether there exists a corresponding renormalon
singularity, has been addressed by several authors, with seemingly
controversial conclusions. At the one-loop order, such a mixing has
indeed been observed when the HQET is regularized on a space--time
lattice \cite{MMS}. Likewise, a ``physical'' definition of a
parameter $\lambda_1(\mu)$ has been suggested, which absorbs certain
$O(\alpha_s)$ corrections appearing in the zero-recoil sum rules for
heavy-quark transitions \cite{BSUV}. This definition is such that
$\mbox{d}\lambda_1(\mu)/\mbox{d}\mu^2\propto \alpha_s(\mu)$,
indicating again a one-loop mixing of the kinetic energy with the
identity. On the other hand, this mixing has not been observed at the
one-loop order in two Lorentz-invariant cutoff regularization schemes
\cite{MNS}, which use a Pauli--Villars regulator or a cutoff on the
virtuality of the gluon in one-loop Feynman diagrams \cite{flow}.
This observation appeared as a puzzle, which was resolved when it was
shown that the mixing between the kinetic energy and the identity is
forbidden at the one-loop order in all regularization schemes with a
Lorentz-invariant UV cutoff,\footnote{The fact that a mixing at the
one-loop order was observed in Refs.~\protect\cite{BSUV,MMS} is a
consequence of the explicit breaking of Lorentz invariance by the
regularization schemes adopted in these calculations.}
but that in general there is no symmetry that protects the matrix
elements of the kinetic operator from quadratic divergences, and so a
mixing with the identity occurs from the two-loop order on
\cite{explo}.

\begin{figure}
\epsfxsize=7.5cm
\epsffile{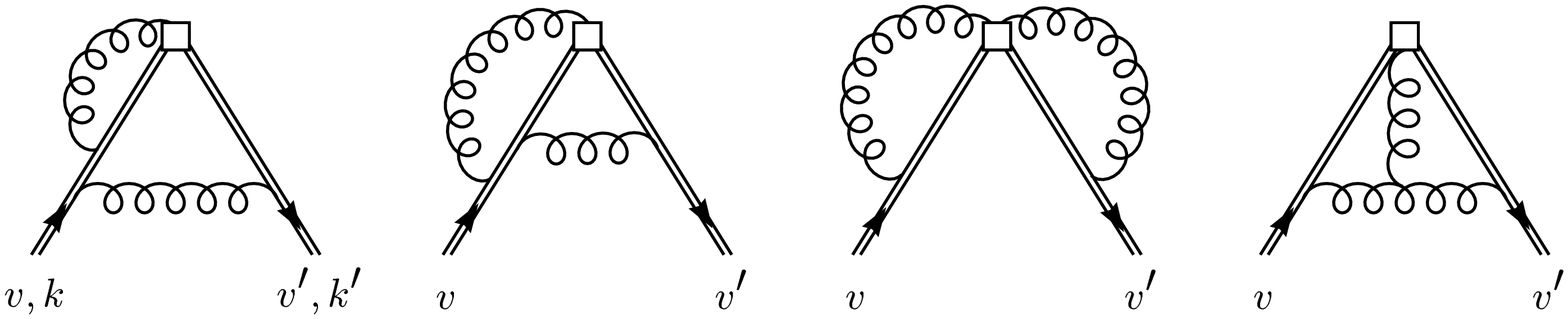}
\vspace{-0.5cm}
\caption{Two-loop diagrams contributing to the mixing of the
chromo-electric operator with lower-dimensional operators. Not shown
are two copies of the first two diagrams with the ``outer'' gluon
attached to the other heavy-quark line.}
\label{fig:2loop}
\end{figure}

In understanding this situation, the virial theorem of the HQET is of
great help. This theorem provides a relation between the kinetic
energy of a heavy quark inside a hadron and its chromo-electric
interactions with gluons \cite{virial}. For the $B$ meson, it can be
written in the form
\begin{eqnarray}
   \lim_{v'\to v}&&\!\frac{\langle B(v')|\,
    \bar h_{v'} v_\mu v'_\nu\,ig_s G^{\mu\nu} h_v\,|B(v)\rangle}
    {(v\cdot v')^2-1} \nonumber\\
   &&= -\frac 13\,\langle B|\,O_{\rm kin}\,|B\rangle \,.
\label{virial}
\end{eqnarray}
Note that in the rest frame of the initial or the final meson, the
operator appearing on the left-hand side of this relations only
contains the chromo-electric field $E_c^i=-G^{0i}$. The virial
theorem is a most useful relation. Not only is it intuitive,
generalizing a well-known concept of classical physics; it also helps
in understanding better the properties of the kinetic operator. For
instance, it has been used to estimate the hadronic parameter
$\lambda_1$ using QCD sum rules \cite{l1sr}. More important, however,
is the fact that the virial theorem can be employed to analyse the
properties of the kinetic operator under renormalization
\cite{explo}. (Parenthetically, we note that this theorem has also
been used in the calculation of the two-loop anomalous dimension of
the chromo-magnetic operator \cite{ABM,Amor}.) In the limit $v'\to
v$, the properties of the two operators $O_{\rm kin}$ and $O_{\rm
el}^{\mu\nu}=\bar h_{v'}ig_s G^{\mu\nu} h_v$ are related to each
other. The chromo-electric operator $O_{\rm el}^{\mu\nu}$ can mix
with the lower-dimensional operator $(v^\mu v'^\nu - v^\nu
v'^\mu)\,\bar h_{v'} h_v$. However, in any regularization scheme with
a Lorentz-invariant ultraviolet cutoff, such a mixing can only come
from Feynman diagrams involving gluons attached to both heavy-quark
lines, since otherwise there is no way to get the factor $(v^\mu
v'^\nu - v^\nu v'^\mu)$. Such diagrams appear first at the two-loop
order, as shown in Fig.~\ref{fig:2loop}. The virial theorem thus
provides for a simple explanation of the fact that the mixing of the
kinetic operator with the identity was not observed at the one-loop
order in Lorentz-invariant regularization schemes. On the other hand,
an explicit calculation confirms that the mixing is present at the
two-loop order, and thus there is indeed a quadratic divergence (and
thus a renormalon problem) affecting the matrix elements of the
kinetic operator \cite{explo}.

At the end of this discussion, we stress that the ``renormalon
ambiguities'' are not a conceptual problem for the heavy-quark
expansion. It can be shown quite generally that these ambiguities
cancel in all predictions for physical quantities
\cite{Chris}--\cite{LMS}. The way the cancellations occur is
intricate, however. The generic structure of the heavy-quark
expansion for an observable is of the form:
\begin{equation}
   \mbox{Observable} \sim C[\alpha_s(m_Q)]\,\bigg( 1
   + {\Lambda\over m_Q} + \dots \bigg) \,,
\end{equation}
where $C[\alpha_s(m_Q)]$ represents a perturbative coefficient
function, and $\Lambda$ is a dimensionful non-perturbative parameter.
The truncation of the perturbation series defining the coefficient
function leads to an arbitrariness of order $\Lambda_{\rm QCD}/m_Q$,
which cancels against a corresponding arbitrariness of order
$\Lambda_{\rm QCD}$ in the definition of the non-perturbative
parameter $\Lambda$. Thus, only when the short- and long-distance
contributions are combined in the heavy-quark expansion, an
unambiguous result is obtained.

\section{INCLUSIVE B DECAY RATES}

Inclusive decay rates determine the probability of the decay of a
particle into the sum of all possible final states with a given set
of global quantum numbers. An example is provided by the inclusive
semileptonic decay rate of the $B$ meson, $\Gamma(B\to X_c\,\ell\,
\bar\nu)$, where the final state consists of a lepton--neutrino pair
accompanied by any number of hadrons with total charm-quark number
$n_c=1$. Here we shall discuss the theoretical description of
inclusive decays of $B$ mesons (an analogous description holds for
all hadrons containing a heavy quark) \cite{Chay}--\cite{MNTM}. From
the theoretical point of view, such decays have two advantages:
first, bound-state effects related to the initial state (such as the
``Fermi motion'' of the heavy quark inside the hadron
\cite{shape,Fermi}) can be accounted for in a systematic way using
the heavy-quark expansion; secondly, the fact that the final state
consists of a sum over many hadronic channels eliminates bound-state
effects related to the properties of individual hadrons. This second
feature is based on a hypothesis known as quark--hadron duality,
which is an important concept in QCD phenomenology. The assumption of
duality is that cross sections and decay rates, which are defined in
the physical region (i.e.\ the region of time-like momenta), are
calculable in QCD after a ``smearing'' or ``averaging'' procedure has
been applied \cite{PQW}. In semileptonic decays, it is the
integration over the lepton and neutrino phase space that provides a
``smearing'' over the invariant hadronic mass of the final state
(so-called global duality). For non-leptonic decays, on the other
hand, the total hadronic mass is fixed, and it is only the fact that
one sums over many hadronic states that provides an ``averaging''
(so-called local duality). Clearly, local duality is a stronger
assumption than global duality. It is important to stress that
quark--hadron duality cannot yet be derived from first principles,
although it is a necessary assumption for many applications of QCD.
The validity of global duality has been tested experimentally using
data on hadronic $\tau$ decays \cite{Maria}. Some more formal
attempts to address the problem of quark--hadron duality can be found
in Refs.~\cite{Shifm,Boyd}.

Using the optical theorem, the inclusive decay width of $B$ mesons
can be written in the form
\begin{equation}\label{ImT}
   \Gamma(B\to X) = \frac{1}{m_B}\,\mbox{Im}\,
   \langle B|\,{\bf T}\,|B\rangle \,,
\end{equation}
where the transition operator ${\bf T}$ is given by
\begin{equation}
   {\bf T} = i\!\int{\rm d}^4x\,T\{\,
   {\cal L}_{\rm eff}(x),{\cal L}_{\rm eff}(0)\,\} \,.
\end{equation}
In fact, inserting a complete set of states inside the time-ordered
product, we recover the standard expression
\begin{eqnarray}
   \Gamma(B\to X) &=& \frac{1}{2 m_B}\,\sum_X\,
    (2\pi)^4\,\delta^4(p_B-p_X) \nonumber\\
   &&\times |\langle X|\,{\cal L}_{\rm eff}\,|B\rangle|^2
\end{eqnarray}
for the decay rate. Here ${\cal L}_{\rm eff}$ is the effective weak
Lagrangian corrected for short-distance effects arising from the
exchange of gluons with virtualities between $m_W$ and $m_b$
\cite{AltM}--\cite{cpcm5}. If some quantum numbers of the final
states $X$ are specified, the sum over intermediate states is to be
restricted appropriately. In the case of the inclusive semileptonic
decay rate, for instance, the sum would include only those states $X$
containing a lepton--neutrino pair.

\begin{figure}
\epsfxsize=7.5cm
\epsffile{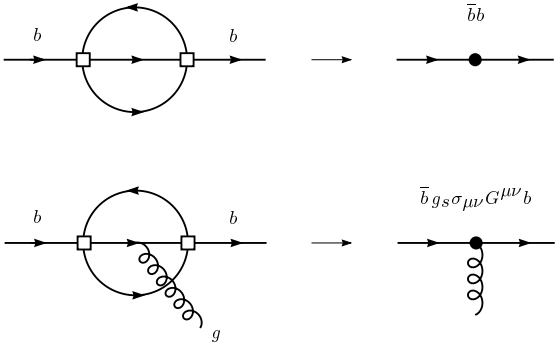}
\vspace{-0.5cm}
\caption{Perturbative contributions to the transition operator ${\bf
T}$ (left), and the corresponding operators in the OPE (right). The
open squares represent a four-fermion interaction of the effective
Lagrangian ${\cal L}_{\rm eff}$, while the black circles represent
local operators in the OPE.}
\label{fig:Toper}
\end{figure}

In perturbation theory, some contributions to the transition operator
are given by the two-loop diagrams shown on the left-hand side in
Fig.~\ref{fig:Toper}. Because of the large mass of the $b$ quark, the
momenta flowing through the internal propagator lines are large. It
is thus possible to construct an OPE for the transition operator, in
which ${\bf T}$ is represented as a series of local operators
containing the heavy-quark fields. The operator with the lowest
dimension, $d=3$, is $\bar b b$. It arises by contracting the
internal lines of the first diagram. The only gauge-invariant
operator with dimension 4 is $\bar b\,i\rlap{\,/}D\,b$; however, the
equations of motion imply that between physical states this operator
can be replaced by $m_b\bar b b$. The first operator that is
different from $\bar b b$ has dimension 5 and contains the gluon
field. It is given by $\bar b\,g_s\sigma_{\mu\nu} G^{\mu\nu} b$. This
operator arises from diagrams in which a gluon is emitted from one of
the internal lines, such as the second diagram shown in
Fig.~\ref{fig:Toper}. For dimensional reasons, the matrix elements of
such higher-dimensional operators are suppressed by inverse powers of
the heavy-quark mass.

In the next step, the hadronic matrix elements of the local operators
in the OPE are expanded in powers of $1/m_b$, using the technology of
the HQET. The result is \cite{FaNe,MaWe,Adam}
\begin{eqnarray}
   \frac{\langle B|\,\bar b b\,|B\rangle}{2 m_B}
   = 1 + \frac{\lambda_1+3\lambda_2}{2 m_b^2} + O(1/m_b^3) \,,
    \nonumber\\
   \frac{\langle B|\,\bar b\,g_s\sigma_{\mu\nu} G^{\mu\nu} b\,
         |B\rangle}{2 m_B} = 6\lambda_2 + O(1/m_b) \,.
\end{eqnarray}
Thus, any inclusive decay rate can be written in the form
\cite{Bigi}--\cite{MaWe}
\begin{eqnarray}\label{generic}
   \Gamma(B\to X_f) &=& \frac{G_F^2 m_b^5}{192\pi^3}\,
    \Bigg\{ c_3^f \left( 1 + \frac{\lambda_1+3\lambda_2}{2 m_b^2}
    \right) \nonumber\\
   &&\mbox{}+ c_5^f\,\frac{\lambda_2}{m_b^2} + O(1/m_b^3) \Bigg\} \,,
\end{eqnarray}
where the prefactor arises naturally from the loop integrations, and
$c_n^f$ are calculable coefficient functions (which also contain the
relevant CKM matrix elements) depending on the quantum numbers $f$ of
the final state. It is instructive to understand the appearance of
the kinetic-energy contribution $\lambda_1/2 m_b^2$. This is nothing
but the field-theory analogue of the Lorentz factor $(1-{\bf
v}_b^2)^{1/2}\simeq 1-{\bf k}^2/2 m_b^2$, in accordance with the fact
that the lifetime, $\tau=1/\Gamma$, for a moving particle (the $b$
quark) increases due to time dilation.

The main result of the heavy-quark expansion for inclusive decay
rates is the observation that the free quark decay (i.e.\ the parton
model) provides the first term in a systematic $1/m_b$ expansion
\cite{Chay}. For dimensional reasons, the corresponding rate is
proportional to the fifth power of the $b$-quark mass. The
non-perturbative corrections, which arise from bound-state effects
inside the $B$ meson, are suppressed by two powers of the heavy-quark
mass, i.e.\ they are of relative order $(\Lambda_{\rm QCD}/m_b)^2$.
Note that the absence of first-order power corrections is a
consequence of the equations of motion, as there is no independent
gauge-invariant operator of dimension 4 that could appear in the OPE.
The fact that bound-state effects in inclusive decays are strongly
suppressed explains a posteriori the success of the parton model in
describing such processes \cite{ACCMM,Pasch}.

The hadronic parameters $\lambda_1$ and $\lambda_2$ appearing in the
heavy-quark expansion (\ref{generic}) have been discussed in the
previous section. For a given inclusive decay channel, what remains
to be calculated is the coefficient functions $c_n^f$. This can be
done using perturbation theory. We shall now discuss three important
applications of this general formalism.

\boldmath
\subsection{Determination of $|V_{cb}|$ from Inclusive Semileptonic
Decays}
\unboldmath

The extraction of $|V_{cb}|$ from the inclusive semileptonic decay
rate of $B$ mesons is based on the general expression
(\ref{generic}), with the short-distance coefficients
\cite{Bigi}--\cite{MaWe}
\begin{eqnarray}
   c_3^{\rm SL} &=& |V_{cb}|^2 \Big[ 1 - 8 x^2 + 8 x^6 - x^8
    - 12 x^4\ln x^2 \nonumber\\
   &&\qquad \mbox{}+ O(\alpha_s) \Big] \,, \nonumber\\
   c_5^{\rm SL} &=& -6 |V_{cb}|^2 (1-x^2)^4 \,.
\end{eqnarray}
Here $x=m_c/m_b$, and $m_b$ and $m_c$ are the pole masses of the $b$
and $c$ quarks, defined to a given order in perturbation theory
\cite{Tarr}. The $O(\alpha_s)$ terms in $c_3^{\rm SL}$ are known
exactly \cite{fgrefs}, while only partial calculations of
higher-order corrections exist \cite{LSW,BaBB}. The main sources of
theoretical uncertainties are the dependence on the heavy-quark
masses, unknown higher-order perturbative corrections, and the
assumption of global quark--hadron duality. A conservative estimate
of the total theoretical error on the extracted value of $|V_{cb}|$
is $\delta|V_{cb}|/|V_{cb}|\approx 10\%$ \cite{Beijing}. Taking the
result of Ball et al.\ \cite{BaBB} for the central value, and using
$\tau_B=(1.60\pm 0.03)\,$ps for the average $B$-meson lifetime
\cite{Rich}, we find
\begin{eqnarray}
   |V_{cb}| &=& (0.040\pm 0.004)\,\bigg( {B_{\rm SL}\over 10.8\%}
    \bigg)^{1/2} \nonumber\\
   &=& (40\pm 1_{\rm exp}\pm 4_{\rm th}) \times 10^{-3} \,.
\end{eqnarray}
In the last step, we have used $B_{\rm SL}=(10.8\pm 0.5)\%$ for the
semileptonic branching ratio of $B$ mesons (see below). The value of
$|V_{cb}|$ extracted from the inclusive semileptonic width is in
excellent agreement with that obtained from the analysis of the
exclusive decay $B\to D^*\ell\,\bar\nu$ using heavy-quark symmetry
\cite{Vcb}--\cite{OPALVcb}. This agreement is gratifying given the
differences of the methods used, and it provides an indirect test of
global quark--hadron duality. Combining the two measurements gives
$|V_{cb}| = 0.039\pm 0.002$ \cite{Beijing}. After $V_{ud}$ and
$V_{us}$, this is now the third-best known entry in the
Cabibbo--Kobayashi--Maskawa (CKM) matrix.

\boldmath
\subsection{Semileptonic Branching Ratio for Decays into $\tau$
Leptons}
\unboldmath

Semileptonic decays of $B$ mesons into $\tau$ leptons are of
particular importance, since they are sensitive probes of physics
beyond the Standard Model \cite{Zoltau}. From the theoretical point
of view, the ratio of the semileptonic rates (or branching ratios)
into $\tau$ leptons and electrons can be calculated reliably. This
ratio is independent of the factor $m_b^5$, the hadronic parameter
$\lambda_1$, and CKM matrix elements. To order $1/m_b^2$, one finds
\cite{incltau,Balk,Koyr}
\begin{eqnarray}
   \frac{B(B\to X\,\tau\,\bar\nu_\tau)}{B(B\to X\,e\,\bar\nu_e)}
   &=& f(x_c,x_\tau) + \frac{\lambda_2}{m_b^2}\,g(x_c,x_\tau)
    \nonumber\\
   &=& 0.22\pm 0.02 \,,
\label{Rtau}
\end{eqnarray}
where $f$ and $g$ are calculable coefficient functions depending on
the mass ratios $x_c=m_c/m_b$ and $x_\tau=m_\tau/m_b$, as well as on
$\alpha_s(m_b)$. Two new measurements of the semileptonic branching
ratio of $b$ quarks, $B(b\to X\,\tau\,\bar\nu_\tau)$, have been
reported by the ALEPH and OPAL Collaborations at LEP
\cite{Btau1,Btau2}. The weighted average is $(2.68\pm 0.28)\%$.
Normalizing this result to the LEP average value $B(b\to
X\,e\,\bar\nu_e)=(10.95\pm 0.32)\%$ \cite{Rich}, we obtain
\begin{equation}
   {B(b\to X\,\tau\,\bar\nu_\tau)\over
    B(b\to X\,e\,\bar\nu_\tau)} = 0.245\pm 0.027 \,,
\end{equation}
in good agreement with the theoretical prediction (\ref{Rtau}) for
$B$ mesons.

\subsection{Semileptonic Branching Ratio and Charm Counting}

The semileptonic branching ratio of $B$ mesons is defined as
\begin{equation}
   B_{\rm SL} = {\Gamma(B\to X\,e\,\bar\nu)\over
   \sum_\ell \Gamma(B\to X\,\ell\,\bar\nu) + \Gamma_{\rm had}
   + \Gamma_{\rm rare}} \,,
\end{equation}
where $\Gamma_{\rm had}$ and $\Gamma_{\rm rare}$ are the inclusive
rates for hadronic and rare decays, respectively. The main difficulty
in calculating $B_{\rm SL}$ is not in the semileptonic width, but in
the non-leptonic one. As mentioned previously, the calculation of
non-leptonic decay rates in the heavy-quark expansion relies on the
strong assumption of local quark--hadron duality.

Measurements of the semileptonic branching ratio have been performed
by various experimental groups, using both model-dependent and
model-independent analyses. The status of the results is
controversial, as there is a discrepancy between low-energy
measurements performed at the $\Upsilon(4s)$ resonance and
high-energy measurements performed at the $Z^0$ resonance. The
situation has been reviewed recently by Richman \cite{Rich}, whose
numbers we shall use in this section. The average value at low
energies is $B_{\rm SL}=(10.23\pm 0.39)\%$, whereas high-energy
measurements give $B_{\rm SL}(b)=(10.95\pm 0.32)\%$. The label $(b)$
indicates that this value refers not to the $B$ meson, but to a
mixture of $b$ hadrons (approximately 40\% $B^-$, 40\% $B^0$, 12\%
$B_s$, and 8\% $\Lambda_b$). Assuming that the corresponding
semileptonic width $\Gamma_{\rm SL}(b)$ is close to that of $B$
mesons,\footnote{Theoretically, this is expected to be a very good
approximation.}
we can correct for this fact and find $B_{\rm
SL}=(\tau_B/\tau_b)\,B_{\rm SL}(b)=(11.23\pm 0.34)\%$, where
$\tau_b=(1.56\pm 0.03)\,$ps is the average lifetime corresponding to
the above mixture of $b$ hadrons. The discrepancy between the low-
and high-energy measurements of the semileptonic branching ratio is
therefore larger than three standard deviations. If we take the
average and inflate the error to account for this fact, we obtain
\begin{equation}\label{Bslval}
   B_{\rm SL} = (10.80\pm 0.51)\% \,.
\end{equation}
An important aspect in interpreting this result is charm counting,
i.e.\ the measurement of the average number $n_c$ of charm hadrons
produced per $B$ decay. Theoretically, this quantity is given by
\begin{equation}\label{ncdef}
   n_c = 1 + B(B\to X_{c\bar c s'}) - B(B\to X_{{\rm no}\,c}) \,,
\end{equation}
where $B(B\to X_{c\bar c s'})$ is the branching ratio for decays into
final states containing two charm quarks, and $B(B\to
X_{{\rm no}\,c})\approx 0.02$ is the Standard Model branching ratio
for charmless decays \cite{Alta}--\cite{Buch}. The average value
obtained at low energies is $n_c=1.12\pm 0.05$ \cite{Rich}, whereas
high-energy measurements give $n_c=1.23\pm 0.07$ \cite{ALEnc}. The
weighted average is
\begin{equation}
   n_c = 1.16\pm 0.04 \,.
\end{equation}

The naive parton model predicts that $B_{\rm SL}\approx 15\%$ and
$n_c\approx 1.2$; however, it has been known for some time that
perturbative corrections could change these predictions significantly
\cite{Alta}. With the establishment of the heavy-quark expansion, the
non-perturbative corrections to the parton model could be computed,
and their effect turned out to be very small. This led Bigi et al.\
to conclude that values $B_{\rm SL}<12.5\%$ cannot be accommodated by
theory \cite{baff}. Later, Bagan et al.\ have completed the
calculation of the $O(\alpha_s)$ corrections including the effects of
the charm-quark mass, finding that they lower the value of $B_{\rm
SL}$ significantly \cite{BSLnew1}. Their original analysis has
recently been corrected in an erratum. Here we shall present the
results of an independent numerical analysis using the same
theoretical input (for a detailed discussion, see
Ref.~\cite{MNChris}). The semileptonic branching ratio and $n_c$
depend on the quark-mass ratio $m_c/m_b$ and on the ratio $\mu/m_b$,
where $\mu$ is the scale used to renormalize the coupling constant
$\alpha_s(\mu)$ and the Wilson coefficients appearing in the
non-leptonic decay rate. The freedom in choosing the scale $\mu$
reflects our ignorance of higher-order corrections, which are
neglected when the perturbative expansion is truncated at order
$\alpha_s$. We allow the pole masses of the heavy quarks to vary in
the range
\begin{eqnarray}
   m_b &=& (4.8\pm 0.2)~\mbox{GeV} \,, \nonumber\\
   m_b - m_c &=& (3.40\pm 0.06)~\mbox{GeV} \,,
\end{eqnarray}
corresponding to $0.25<m_c/m_b<0.33$. The value of the difference
$m_b-m_c$ is obtained from (\ref{mQdif}) using $\lambda_1=-(0.4\pm
0.2)\,\mbox{GeV}^2$. Non-perturbative effects appearing at order
$1/m_b^2$ in the heavy-quark expansion are described by the single
parameter $\lambda_2$ defined in (\ref{lam2val}); the dependence on
the parameter $\lambda_1$ is the same for all inclusive decay rates
and cancels out in the predictions for $B_{\rm SL}$ and $n_c$. For
the two choices $\mu=m_b$ and $\mu=m_b/2$, we obtain \cite{MNChris}
\begin{eqnarray}
   B_{\rm SL} &=& \cases{
    12.0\pm 1.0 \% ;& $\mu=m_b$, \cr
    10.9\pm 1.0 \% ;& $\mu=m_b/2$, \cr} \nonumber\\
   \phantom{ \bigg[ }
   n_c &=& \cases{
    1.20\mp 0.06 ;& $\mu=m_b$, \cr
    1.21\mp 0.06 ;& $\mu=m_b/2$. \cr}
\end{eqnarray}
The uncertainties in the two quantities, which result from the
variation of $m_c/m_b$ in the range given above, are anticorrelated.
Notice that the semileptonic branching ratio has a stronger scale
dependence than $n_c$. By choosing a low renormalization scale,
values $B_{\rm SL}<12\%$ can easily be accommodated. This is indeed
not unnatural. Using the BLM scale-setting method \cite{BLM}, it has
been estimated that $\mu\gsim 0.32 m_b$ is an appropriate scale to
use in this case \cite{LSW}.

\begin{figure}
\epsfxsize=7.5cm
\epsffile{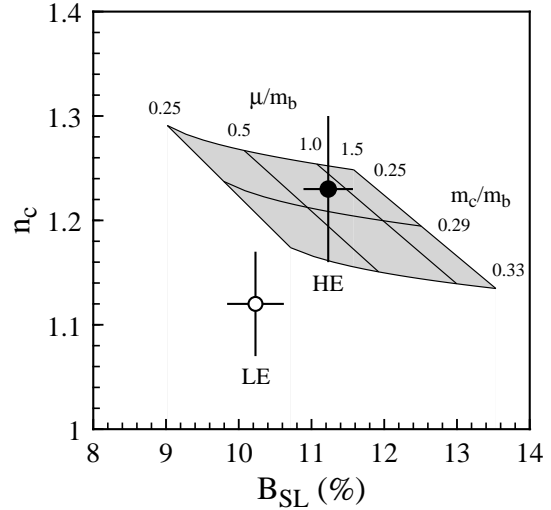}
\vspace{-0.5cm}
\caption{Theoretical prediction for the semileptonic branching ratio
and charm counting as a function of the quark-mass ratio $m_c/m_b$
and the renormalization scale $\mu$. The data points show the average
experimental values obtained in low-energy (LE) and high-energy (HE)
measurements, as discussed in the text.}
\label{fig:BSL}
\end{figure}

The combined theoretical predictions for the semileptonic branching
ratio and charm counting are shown in Fig.~\ref{fig:BSL}. They are
compared with the experimental results obtained from low- and
high-energy measurements. It has been argued that the combination of
a low semileptonic branching ratio and a low value of $n_c$ would
constitute a potential problem for the Standard Model \cite{Buch}.
However, with the new experimental and theoretical numbers, only for
the low-energy measurements a discrepancy remains between theory and
experiment. Note that, with (\ref{ncdef}), our results for $n_c$ can
be used to calculate the branching ratio $B(B\to X_{c\bar c s'})$,
which is accessible to a direct experimental determination. Our
prediction of $(22\pm 6)\%$ for this branching ratio agrees well with
the preliminary result reported by the CLEO Collaboration, which is
$B(B\to X_{c\bar c s'})=(23.9\pm 3.8)\%$ \cite{Hons}.

\section{SUMMARY}

We have presented the theory of inclusive decays of hadrons
containing a heavy quark, and discussed some of its most important
applications to the decays of $B$ mesons: the determination of
$|V_{cb}|$ from inclusive semileptonic decays, semileptonic decays
into $\tau$ leptons, and the semileptonic branching ratio. The
theoretical tools that allow us to perform quantitative calculations
are the heavy-quark symmetry, the heavy-quark effective theory, and
the $1/m_Q$ expansion. In the case of inclusive decay rates, the
non-perturbative information entering the theoretical description is
encoded in two hadronic parameters, $\lambda_1$ and $\lambda_2$ (or
$\mu_\pi^2$ and $\mu_G^2$). We have reviewed the theoretical
understanding of these parameters, concerning both their numerical
values and their properties under renormalization. The parameter
$\lambda_2$, renormalized at the scale $\mu=m_b$, can be extracted
from the spectroscopy of $B$ mesons. The parameter $\lambda_1$, on
the other hand, suffers from a renormalon ambiguity problem and thus
needs a non-perturbative subtraction to be well defined.

\end{document}